\input phyzzx
\def\CAMBRIDGE{\address{Department of Applied Mathematics and
Theoretical Physics\break University of Cambridge\break
Cambridge CB3 9EW, England\break email\ \ p.m.sutcliffe@amtp.cam.ac.uk}}
\def\waveop {\bx80{.6}}
\font\upright=cmu10 scaled\magstep1
\font\sans=cmss12
\def\ss{\sans}
\def\stroke{\vrule height8pt width0.4pt depth-0.1pt}
\def\Ctext{{\rlap{\rlap{C}\kern 3.8pt\stroke}\phantom{C}}}
\def\Rtext{\hbox{\upright\rlap{I}\kern 1.7pt R}}
\def\Ztext{\hbox{\upright\rlap{\ss Z}\kern 2.7pt {\ss Z}}}
\def\Ptext{{\upright\rlap{I}\kern 1.7pt P}}
\def\Ptwotext{\hbox{\upright\rlap{I}\kern 1.7pt P}}
\def\C{\ifmmode{{\hbox\Ctext}}\else\Ctext\fi}
\def\P{\ifmmode{{\hbox\Ptext}}\else\Ptext\fi}
\def\Z{\ifmmode{\vcenter{\Ztext}}\else\Ztext\fi}
\def\R{\ifmmode{\vcenter{\Rtext}}\else\Rtext\fi}
\def\P{\ifmmode{{\hbox\Ptext}}\else\Ptext\fi}
\def\CP{\C\P}

\def\ie{{\sl ie}}
\def\identity{{\upright\rlap{1}\kern 2.0pt 1}}
\def\mbf#1{\setbox0=\hbox{#1}}
\def\bx#1#2#3{{\dimen1=#1pt\advance\dimen1 by #2pt
               \dimen2=#2pt\advance\dimen2 by #3pt
               \dimen3=#1pt\advance\dimen3 by -#3pt\advance\dimen3 by -#3pt
               \dimen4=#3pt\advance\dimen4 by -#2pt\advance\dimen4 by -#1pt
               \vrule height \dimen1 depth -#2pt width #3pt
               \vrule height \dimen2 depth -#2pt width\dimen3
               \hskip\dimen4
               \vrule height \dimen1 depth \dimen4 width\dimen3
               \vrule height \dimen1 depth -#2pt width #3pt}}
\def\.{.\thinspace }

\def\AMALL{M\.F\.Atiyah and
N\.S\.Manton\journal~Phys.~Lett.&222B(89)438.
\nextline M\.F\.Atiyah and
N\.S\.Manton\journal~Comm.~Math.~Phys.&153(93)391.
\nextline N\.S\.Manton, \lq\lq{\sl Skyrme fields and instantons}\rq\rq
\nextline in
Geometry of Low Dimensional Manifolds: 1
(LMS Lecture Notes 150),\nextline eds. S\.K\.Donaldson and
C\.B\.Thomas: Cambridge University Press 1990.}
\def\NSME{N\.S\.Manton\journal~Phys.~Rev.~Lett.&60B(88)1916.}
\def\MFAB{M\.F\.Atiyah and N\.J\.Hitchin,
\lq\lq{\sl~The Geometry and Dynamics of Magnetic\nextline
          Monopoles~\/}\rq\rq\ (Princeton University Press, Princeton,
1988).}
\def\TMSB{T\.M\.Samols\journal~Comm.~Math.~Phys.&145(92)149.}
\def\PJRA{P\.J\.Ruback\journal~Nucl.~Phys.&296B(88)669.}
\def\IABSA{I\.A\.B\.Strachan\journal~J.~Math.~Phys.&33(92)102.}
\def\RSWC{R\.S\.Ward\journal~Phys.~Lett.&158B(85)424.}
\def\RALA{R\.A\.Leese\journal~Nucl.~Phys.&344B(90)33.}
\def\GWGA{G\.W\.Gibbons and
N\.S\.Manton\journal~Nucl.~Phys.&274B(86)183.}
\def\BSA{B\.J\.Schroers\journal~Nucl.~Phys.&367B(91)177.}
\def\PMSA{P\.M\.Sutcliffe\journal~Nonlinearity&4(91)1109.}
\def\SHAH{P\.A\.Shah, \lq\lq{\sl Vortex scattering at  near critical 
coupling}\rq\rq,\nextline Cambridge preprint DAMTP 94-8.}
\def\THSA{T\.H\.R\.Skyrme\journal~Nucl.~Phys.&31(62)556.}
\def\EWB{E\.Witten \journal~Nucl.~Phys.&160B(79)57.}
\def\ANW{G\.S\.Adkins, C\.R\.Nappi and E\.Witten
\journal~Nucl.~Phys.&228B(83)552.}
\def\THSB{T\.H\.R\.Skyrme \journal~Proc.~R.~Soc.&262A(61)237.}
\def\PSKY{J\.K\.Perring and
T\.H\.R\.Skyrme\journal~Nucl.~Phys.&31(62)550.}
\def\PMSD{P\.M\.Sutcliffe\journal~Phys.~Lett.&283B(92)85.}
\def\SW{G\.N\.Stratopoulos and
W\.J\.Zakrzewski\journal~Z.~Phys.~C&59(93)307.}
\def\TH{G\.'t Hooft, {\sl unpublished.}}
\def\CF{E\.F\.Corrigan and D\.B\.Fairlie
\journal~Phys.~Lett.&67(77)69.}
\def\WIL{F\.Wilczek, {\sl\lq\lq Quark confinement and field theory\rq\rq}
\nextline { ed. D\.Stump and D\.Weingarten (John Wiley, New York, 1977).}}
\def\JNR{R\.Jackiw, C\.Nohl and
C\.Rebbi\journal~Phys.~Rev.&15D(77)1642.}
\def\ADHM{M\.F\.Atiyah, V\.G\.Drinfield, N\.J\. Hitchin
and Yu\.I\. Manin\journal~Phys.~Lett.&65A(78)185.}
\def\COFN{E\.F\.Corrigan, D\.I\.Olive, D\.B\.Fairlie and J\.Nuyts
\journal~Nucl.~Phys.&140B(78)31.}
\def\HGOA{A\.Hosaka, S\.M\.Griffies, M\.Oka and R\.D\.Amado
\journal~Phys.~Lett.&251B(90)1.}
\def\HOA{A\.Hosaka, M\.Oka and R\.D\.Amado
\journal~Nucl.~Phys.&530A(91)507.}
\def\PMSF{P\.M\.Sutcliffe\journal~Nucl.~Phys.&393B(93)211.}
\def\PMSH{P\.M\.Sutcliffe\journal~Phys.~Lett.&328B(94)84.}
\def\PMSE{P\.M\.Sutcliffe\journal~Phys.~Lett.&292B(92)104.}
\input epsf
\Tenpoint
\def\boldsig {{\mbf{$\sigma$}}}
\def\mn {${\cal M}_n$ }
\def\ie {{\sl ie }}
\def\cpsm {$\CP^1$\ $\sigma$-model }
\def\cp {$\CP^1$}
\def\eval {\biggr\vert}
\def\waveop {\bx80{.6}}
\Pubnum={DAMTP 94-55}
\date={August 1994}

\titlepage
\title{{\bf Instanton Moduli and Topological Soliton Dynamics}
\footnote \ {To appear in Nuclear Physics B}}
\vskip -1cm
\author{Paul M. Sutcliffe}
\CAMBRIDGE

\abstract
It has been proposed by Atiyah and Manton \REF\rAMALL\AMALL[$\rAMALL$]
 that the 
dynamics of Skyrmions in \R$^{3+1}$ may be approximated by motion on a 
finite dimensional manifold obtained from the moduli space of SU(2)
Yang-Mills instantons in \R$^4$. Motivated by this work we describe 
how similar results exist for other soliton and instanton systems.
We describe in detail two examples for the approximation of the
infinite dimensional dynamics of sine-Gordon solitons in \R$^{1+1}$
by finite dimensional dynamics on a manifold obtained from 
instanton  moduli. In the first example we use the moduli space of 
$\CP^1$ instantons in \R$^2$ and in the second example we use the
moduli space of SU(2) Yang-Mills instantons in \R$^4$. The metric and
potential functions on these manifolds are constructed and the
resulting dynamics is compared with the explicit exact soliton
solutions of the sine-Gordon theory. 
\endpage
\chapter{Introduction}
There are numerous examples in theoretical physics of theories which
possess topological solitons, which are extended structure solutions
with a particle-like interpretation. In (1+1)-dimensions there are models
which are integrable at both the classical and quantum level, thereby
allowing an explicit and exact study of dynamical soliton
interactions. However, in ($d+1$)-dimensions there are no known
integrable models for $d>1$ if one restricts to lagrangian theories
with the physically desirable property of Lorentz covariance. We are
therefore led to the study of non-integrable nonlinear field and gauge
theories, which even at the classical level is a difficult task.
Invariably analytical and/or numerical approximations are required.
For $d>1$ an investigation of classical soliton dynamics through a
numerical simulation of the field equations requires the use of
a large amount of computing power for even a modest study. 
Furthermore, it is unclear how to address the quantum aspects of these
theories from a consideration of the full field theory; which are in
fact often non-renormalizable. It is therefore evident that there
is sufficient motivation for attempting to find analytical
approximation schemes. This paper is concerned with the construction,
implementation and accuracy of such schemes for the situation in which there
are forces between static solitons.

For theories of Bogomolny type there are no forces between static
solitons and the second order nonlinear partial differential
equation which determines the static soliton solutions can be 
reduced to a first order equation. For such theories there is an
analytical approximation scheme which works remarkably well;
namely the geodesic approximation [$\REF\rNSME\NSME\rNSME$].
Here one assumes that the $n$-soliton configuration at any
fixed time may be well approximated by a static $n$-soliton
solution. The only time dependence allowed is therefore in
the dynamics of the finite dimensional $n$-soliton moduli space \mn
(effectively the parameter space of static $n$-soliton solutions).
A lagrangian on \mn is inherited from the field theory lagrangian,
but since all elements of \mn have the same potential energy
the kinetic part of the action completely determines the dynamics. 
It determines a metric on \mn and the dynamics is given by geodesic 
motion on \mn with respect to this metric. The geodesic
approximation has been applied to the study of soliton dynamics
in many theories of Bogomolny type including BPS monopoles
[$\REF\rMFAB\MFAB\rMFAB$],
vortices\def\SRS{\TMSB\nextline\PJRA\nextline\IABSA}
at critical coupling [$\REF\rSRS\SRS\rSRS$] and $\sigma$-model 
lumps\def\WL{\RSWC\nextline\RALA} [$\REF\rWL\WL\rWL$].
For some of these systems numerical simulations of the field theory
equations of motion have been performed and demonstrate the validity
of the geodesic approximation. Quantum soliton interactions
can also be studied by quantization of the finite dimensional system,
which has been studied in great detail for example in the case of BPS 
monopoles\def\GMS{\GWGA\nextline\BSA} [$\REF\rGMS\GMS\rGMS$].

Turning now to the situation in which forces exist between
static solitons ({\sl ie} non-Bogomolny theories) it is clear
that if a truncation to a finite dimensional manifold is to be
performed then a substitute for the moduli space must be found.
In  the special case that the theory may be considered as a
perturbation of a Bogomolny
theory then one may use the moduli space of this associated theory.
The difference now is that there will be both a metric and a potential
function on \mn so that the dynamics will no longer be given by
geodesic motion on \mn but by a more complicated dynamics determined
from the lagrangian on \mn. Such a scheme was applied to a planar
Skyrme-like model \REF\rPMSA\PMSA[$\rPMSA$], considered as a
perturbation of the O(3) $\sigma$-model (a Bogomolny theory), and
found to produce accurate results when compared with numerical
simulations of the field equations. Other theories which are
natural perturbations of Bogomolny theories are, of course, non-BPS
monopoles and vortices which are not at critical coupling. 
Very recently this scheme has been applied to the study of vortex
scattering at  near
critical coupling \REF\rSHAH\SHAH[$\rSHAH$].

In the general case it is more difficult to find a substitute
for the moduli space. One approach is to use the parameter space
of an $n$-soliton configuration obtained by patching together $n$ copies
of a single soliton. This will certainly be an adequate manifold to 
describe solitons which are well separated when inter-soliton
forces are weak. However, it may not provide an adequate description of
solitons which are close together, when the  solitons
become greatly distorted and lose their individual identities.
In particular this patching method is inadequate for the Skyrme model,
where the minimal energy 2-soliton can not be described by a patched
configuration of two single solitons. 
One proposal for the Skyrme model is to use the union of gradient
flow curves descending from a suitable family of static
saddle point solutions [$\rNSME$]. 
Recently Atiyah and Manton
[$\rAMALL$] have shown that an apparently very similar manifold
can be obtained from the moduli space of SU(2) Yang-Mills 
instantons in \R$^4$ by calculating holonomies. They have shown
that it provides a good truncation of the 1-soliton sector and
have investigated the differential and topological structure in
the 2-soliton sector. This connection between instantons and solitons
is still somewhat mysterious, and a clear explanation of the
underlying reason for its existence is still lacking.
There are some technical difficulties 
which together with the complexity of the calculations are responsible
for the lack of a result at present  on the metric and potential functions on
this manifold.

In this paper we describe in detail two simpler examples of an
Atiyah-Manton-like procedure as applied to a truncation of the 
sine-Gordon model in (1+1)-dimensions. We show how to obtain
suitable finite dimensional manifolds from both $\CP^1$ instanton
moduli in \R$^2$ and SU(2) Yang-Mills instanton moduli in \R$^4$.
By calculating the metric and potential functions on these manifolds
we obtain an approximation to the dynamics of solitons in the one and two
soliton sectors. These results are compared with the exact sine-Gordon
soliton solutions. 
We find that the method works well; however,
an unexpected discovery is that a constraint must be placed on the 
orientation of the instantons in order to avoid a spurious bound state
in the two soliton sector.

The plan of the paper is as follows. In section 2 we briefly 
review the Skyrme model and describe how we view the sine-Gordon
theory as its lower dimensional analogue. In section 3 we 
compute sine-Gordon kink fields from $\CP^1$ instantons and discuss
the finite dimensional manifold this produces. Section 4 describes
an alternative possibility to the previous section where we
use Yang-Mills instantons. Section 5 is devoted to an analysis of the
dynamics generated on the finite dimensional manifolds obtained
in the previous two sections. 
Finally in section 6
we make some concluding remarks and observations.

\chapter{Topological Solitons in the Skyrme and Sine-Gordon theories}
The Skyrme model is a nonlinear field theory which provides an
effective
description of low energy hadron physics. Originally introduced by
Skyrme [$\REF\rTHSA\THSA\rTHSA$] this model
 has been the subject of a great deal
of recent study following the observations of Witten [$\REF\rEWB\EWB\rEWB$].
The field of the model is an SU(2) matrix $U$
with lagrangian density given by 
$$
{\cal L} =-{1\over 2}{\rm Tr}(\partial_\mu U\partial^\mu U^{-1})
+{1\over 16}{\rm Tr}([U^{-1}\partial_\mu U,U^{-1}\partial_\nu U]
[U^{-1}\partial^\mu U,U^{-1}\partial^\nu U])
+{m^2\over 2}{\rm Tr}(U+U^\dagger-2)\eqn\lagsky
$$
where $x^\mu =(t,{\bf x})$, $\mu =0,1,2,3$ are the spacetime coordinates in
Minkowski space with metric $\eta^{\mu\nu}={\rm diag}(-1,1,1,1)$
and Tr denotes trace. We have scaled away some phenomenological
constants by a suitable choice of energy and length units, and $m$
is a small constant equal to the pion mass. For the purposes of our
discussion we can neglect the pion mass term and will therefore
consider the Skyrme model with $m=0$ unless otherwise stated.

In order to consider configurations with finite energy the boundary
condition
$$U({\bf x})\rightarrow\identity \hskip 2cm \rm{as} \hskip 1cm
\vert{\bf x}\vert \rightarrow \infty\eqn\bcs$$
is imposed. This effectively compactifies space from \R$^3$ to $S^3$,
so that at a fixed time, $U$ is a map $U({\bf x}):S^3\rightarrow$SU(2).
Since $S^3$ is the group manifold of SU(2) the relevant identity is
the homotopy group relation
$$\Pi_3(S^3)=\Z\eqn\hom$$
which implies that to each configuration there may be associated an
integer $B$, which is conserved and represents the winding number of
the
field as a map from space to the target manifold. It is known as the
topological charge and counts the number of solitons in the
configuration. It is physically identified as the baryon number so
that a soliton is interpreted as a baryon. Explicitly $B$ is given by
$$B={1\over 24\pi^2}\int\epsilon_{ijk}{\rm Tr}(
(U^{-1}\partial_i U)(U^{-1}\partial_j U)(U^{-1}\partial_k U)) d^3{\bf
x}\eqn\topchs$$
where indices range over the values 1,2,3.
There is a lower bound on the energy of a given configuration in terms of
the soliton number {\sl ie} $E\ge 12\pi^2\vert B\vert$.

The static one-soliton solution ({\sl ie} $B=1$) is known as the 
Skyrmion and has the hedgehog form
$$U=\exp(if(r)\hat{\bf x}.\boldsig) 
\hskip 1cm {\rm where} \hskip 1cm \hat{\bf x}={{\bf x}\over r}, 
\hskip 1cm r=\vert{\bf x}\vert
\eqn\hh$$
and $\boldsig$ denote the Pauli matrices. $f(r)$ is a profile function,
with boundary conditions $f(0)=\pi$ , $f(\infty)=0$,
which has to be determined numerically [$\REF\rANW\ANW\rANW$]. This Skyrmion
is situated at the origin but may be moved to any position by a
translation. The orientation may also be changed by conjugation of $U$
by a fixed element of SU(2). The Skyrmion has energy $E_1=1.23\times
12\pi^2$ and so does not saturate the topological lower bound on the
energy. This allows the possibility of a bound state of two solitons
in the $B=2$ sector, and indeed a bound state exists, although it
is not of the hedgehog form. The complicated nature of the 2-soliton
sector and the forces between individual solitons make the study of
soliton dynamics in the Skyrme model a difficult problem. This is the
reason that the approximation scheme proposed in [$\rAMALL$] is both
useful and interesting. 

We now describe the sine-Gordon model in a way that illustrates its
role in this paper, as a lower dimensional analogue of the Skyrme
model. In fact Skyrme himself
[$\REF\rTHSB\THSB\rTHSB$] proposed the sine-Gordon
model as a toy model for a nonlinear meson field theory with the
baryon interpretation of solitons {\ie} the forerunner of the Skyrme
model. Let us consider the Skyrme lagrangian $\lagsky$ (including
the pion mass term) not in 
(3+1)-dimensions with an SU(2)-valued field, but in (1+1)-dimensions
with a U(1)-valued field. We can then write 
$$U=e^{i\phi}\eqn\utophi$$
where $\phi$ is a real-valued field, upon which $\lagsky$ becomes
(after dividing through by the constant 4 to obtain the standard
normalization)
$${\cal L}_{sg}={1\over 8}\partial_\mu\phi\partial^\mu\phi
+{m^2\over 4}(\cos\phi -1)\eqn\lagsg$$
where now $x^\mu=(t,x)$, $\mu=0,1$. This is of course the sine-Gordon
lagrangian. Note that in three dimensional space it is not
 necessary to include the
potential term (pion mass term) in order to have a finite size soliton
solution, although the term quadratic in derivatives (the 
Skyrme term) is required. In contrast in one dimensional space the
Skyrme term is not required (the Skyrme term contribution has vanished
in $\lagsg$) but the potential term must be present. This can easily
be seen as a consequence of the behaviour of the sigma model
term (the first term) under a scale transformation of the space
coordinates. In the intermediate case of two dimensional space both
a Skyrme term and a potential term are required, and this produces the
planar Skyrme-like model  which we mentioned briefly in the introduction.
Since we require $m\neq 0$ in the sine-Gordon lagrangian we can choose
length units to set $m=1$ in $\lagsg$ without loss of generality.

The sine-Gordon equation which follows from $\lagsg$ is
$$\partial_\mu\phi\partial^\mu\phi +\sin\phi=0.\eqn\eomsg$$
The soliton solutions of the sine-Gordon equation (which are known as
kinks) have finite energy so that the field $\phi(x)$ at fixed time
must tend to an integer multiple of $2\pi$ as $x\rightarrow\pm\infty$.
As with the Skyrme model there is therefore an integer-valued conserved
topological charge $B_{sg}$, given by
$$B_{sg}={1\over 2\pi}\int_{-\infty}^{+\infty}\partial_x\phi\ dx.
\eqn\topchsg$$
Again this counts the number of solitons (kinks) in a given
configuration and we regard it as the analogue of the baryon number $B$
of the Skyrme model. There is also a lower bound on the energy of 
a configuration in terms of the soliton number {\sl ie.} 
$E\ge 2\pi\vert B_{sg}\vert $.

The static one-kink solution is given by
$$\phi=4{\rm arctan}(e^{x-a})\eqn\okink$$
where $a$ is the position of the kink. This solution saturates the 
energy bound so that it has energy $E=2$. Note that this is an
important difference between the sine-Gordon model and the Skyrme
model
since it implies that there can be no soliton bound states in the 
sine-Gordon theory (although there are soliton-antisoliton bound
states known as breathers). Another difference between the two models,
which will be important in the next section, is that there is no 
internal phase or orientation for a kink, in contrast to the already
mentioned orientation of the Skyrmion.

There are no static two-kink solutions to the sine-Gordon equation,
but the integrability of the theory allows the explicit construction
of the exact two-kink solution where both kinks have arbitrary
velocity (but at least one non-zero). The solution
$$\phi=4{\rm arctan}\bigl({u\sinh (\gamma x)\over\cosh (\gamma ut)}
\bigr)\eqn\tkink$$
where $\gamma =(1-u^2)^{-{1\over 2}}$, $0<u<1$,
 was first found by
Perring and Skyrme [$\REF\rPSKY\PSKY\rPSKY$] and describes two kinks which
approach along the $x$-axis with velocity $u$, scatter elastically at 
$t=0$, and emerge from the interaction with a phase shift given by
$$\delta={2\log(u^{-1})\over\gamma}.\eqn\psex$$ 
This 2-kink solution has energy $E=4\gamma$.

In the following section we briefly review the principal idea of the
Atiyah-Manton approach to a finite dimensional truncation of the 
Skyrme theory, and then we construct in detail an analogous finite
dimensional truncation of the sine-Gordon theory.

\chapter{Kinks from $\CP^1$ Instantons}
Let $A_\mu$ be the $su(2)$-valued Yang-Mills gauge potential for an
instanton in
four-dimensional euclidean space with coordinates $x_\mu=({\bf x},x_4)$,
$\mu=1,2,3,4$. Note that the euclidean time $x_4$ in the instanton
theory is quite distinct from (and not to be confused with) the 
\lq\lq physical time\rq\rq\  $t$ of the soliton theories introduced in
the previous section. The main idea of the Atiyah-Manton
scheme [$\rAMALL$] is to construct the holonomy
$$U({\bf x})={\cal P}\exp\int_{-\infty}^{+\infty}-A_4({\bf x},x_4)dx_4
\eqn\amhol$$
where {$\cal P$} denotes path ordering.

The fundamental topological result is that if we consider this
holonomy as a Skyrme field then it has baryon number $B=n$, where $n$
is the topological charge (\ie instanton number) of the gauge
potential $A_\mu$. The moduli space of charge $n$ SU(2) Yang-Mills instantons
has dimension $8n-3$, and this generates a manifold of Skyrme fields
with dimension $8n-1$ (after including a choice of gauge at infinity).
It turns out that this manifold appears to be  a good truncation of the Skyrme
model.  Here we will only mention the example for $n=1$, where a 
1-instanton positioned at the origin with scale $\lambda$ generates
a Skyrmion of the hedgehog form $\hh$ with a profile function
$$f(r)=\pi(1-{\lambda r\over\sqrt{1+\lambda^2 r^2}}).\eqn\aspf$$
This Skyrme field has minimum energy when $\lambda= 0.69$
at which it takes the value $\tilde E_1=1.24\times 12\pi^2$, which
is within 1\% of the numerically determined value $E_1$.

There is a lower dimensional analogue of this procedure which was
briefly introduced by the present author in \REF\rPMSD\PMSD
[$\rPMSD$]. Here we describe this result in more detail and use it 
to construct finite dimensional manifolds which we use in section 5 to
study soliton dynamics.

We use the gauge field formulation of the \cpsm in two-dimensional
euclidean space. It is defined in terms of a 2-component column
 vector $Z$, which is a function
of the euclidean spacetime coordinates $x^\mu=(x^1,x^2)$, and is
constrained to satisfy the condition            
$$ Z^\dagger Z= 1.\eqn\con$$
The action density  has a U(1) gauge symmetry and is given by
$$ {\cal L}={\rm Tr}(D_\mu Z)^\dagger(D^\mu Z) \eqn\actcp$$
where Tr denotes trace and $D_\mu$ are the covariant derivatives
$$D_\mu=\partial_\mu - A_\mu \eqn\cov$$
with the composite gauge fields being purely imaginary and defined by
$$ A_\mu=Z^\dagger\partial_\mu Z.\eqn\gf$$ 
The equation of motion resulting from $\actcp$ is
$$[\partial_\mu \partial^\mu \P,\P]=0 \eqn\cpeom$$
where $\P$ is the one-dimensional hermitian projector
$$\P=ZZ^\dagger .\eqn\proj$$
Instantons are finite action solutions to $\cpeom$
and are most easily written using the parametrization
$$Z={1\over \sqrt{1+\vert W \vert ^2}}\left(
\matrix{1\cr W\cr}\right) 
\eqn\wtoz$$
where they are given by 
$W$ a rational function of $x_+=x_1+ix_2$.

The target manifold is \cp , which is isometric to $S^2$,
so that due to the homotopy relation
$$\Pi_2(S^2)=\Z \eqn\homb$$
each finite action
field configuration has an associated integer
winding number. This winding number is the 
topological charge (instanton number) of the
configuration and is given by
$$n={1 \over \pi} \int {{\rm Im}(\partial_1 W \partial_2 \bar W)
\over (1+\vert W \vert ^2)^2} d^2x
 \eqn\topa$$
where Im denotes the imaginary part.
For instanton fields the degree of the rational function $W(x_+)$ is
equal to the instanton number $n$.

Following $\amhol$ we construct the holonomy
$$ U(x_1)=(-1)^n\exp \bigl(\int_{-\infty}^{+\infty} A_2(x_1,x_2) dx_2 \bigr)
\eqn\myhol$$
which is a U(1)-valued field. Note that there is no path ordering
required here since we are dealing with an abelian gauge theory.
The prefactor in $\myhol$ may be interpretated (see below) as the
 extra holonomy
along a large semicircle at infinity in the $(x_1,x_2)$ plane which 
connects the points $x_2=+\infty$ and $x_2=-\infty$ and arises from
the fact that the instanton may be in a singular gauge at infinity.
In fact the holonomy $\amhol$ also requires a prefactor of -1 if the
instanton is in a singular gauge at infinity [$\rAMALL$]. 

We now make use of our interpretation of the sine-Gordon model as a U(1)
Skyrme model through the relation $\utophi$ \ie from the holonomy
$\myhol$ we construct the sine-Gordon field through the definition
$$e^{i\phi(x)}=U(x).\eqn\utophib$$
The fundamental topological result is that the sine-Gordon field 
defined in this way has a kink number equal to the topological
charge of the instanton gauge potential $A_\mu$ \ie $B_{sg}=n$. 
This can be proved using the following steps. Combining 
$\myhol$ with $\utophib$ we have that
$$\phi(x)=n({\rm mod}2)\pi
-i\int_{-\infty}^{+\infty} A_2 dx_2\eval _{x_1=x} \eqn\atophia$$
Let $L$ denote the line which joins the point $(x_1,-\infty)$ to
the point $(x_1,+\infty)$, and $C$ be the large semicircle at infinity
which joins these points with opposite sense of direction.
Then we can write $\atophia$ as
$$
\phi(x)=n({\rm mod}2)\pi
+i \int_C A_\mu dx_\mu\eval _{x_1=x}
-i \int_{L+C} A_\mu dx_\mu\eval _{x_1=x}
\eqn\atophib$$ 
The second term may be evaluated to give $-n\pi$ which cancels exactly
the first term (since $\phi$ takes values in $S^1$ we can
freely add integer multiples of $2\pi$).
The remaining term is an integral over a closed contour
so we can use Stokes\rq\ theorem to write it as
an area integral. In terms of $W$ the  result is
$$\phi(x)=\int_{-\infty}^{x}dx_1
\int_{-\infty}^{+\infty}dx_2 
\ {2{\rm Im}(\partial_1 W \partial_2 \bar W)
\over (1+\vert W \vert ^2)^2}.\eqn\wtopc$$
Comparing with equation $\topa$ we see that $\phi(x)$ is
exactly  $2\pi$ times the instanton topological charge
contained in the region of the $(x_1,x_2)$ plane given by $x_1<x$.
Hence $\phi(-\infty)=0$ and $\phi(\infty)=2\pi n$, and we have the final
result
$$B_{sg}={1\over 2\pi}\int_{-\infty}^{+\infty}\partial_x\phi\ dx
={1\over 2\pi}(\phi(\infty)-\phi(-\infty))=n\eqn\done$$

The moduli space of charge $n$ \cp\ instantons has dimension $4n-1$.
Since the holonomy $\myhol$ is invariant under translations of $x_2$ the
above procedure produces kink fields which depend on a manifold,
which we shall call ${\cal M}_n^{cp}$, that has dimension $4n-2$.

Let us consider the case $n=1$. Then we may take $W$ to be given
by $$W=\lambda(x_+-a)\eqn\wone$$
where $\lambda$ is real and positive and is the scale of
the instanton, and $a$ may be taken to be real due to the above
mentioned invariance. This generates the 1-kink field
$$\phi=\phi_1^{cp}(x,\lambda,a)=
\pi(1+{\lambda(x-a)\over\sqrt{1+\lambda^2 (x-a)^2}}).\eqn\cpo$$
We therefore have that ${\cal M}_1^{cp}={\rm\R}^+\times{\rm\R}$, with the
interpretation of the coordinates as a scale and position for the
kink. The energy of $\cpo$ is obviously independent of $a$, and
is minimized when $\lambda=\lambda_{cp}=0.695$
 at which it takes the value
$E=M_{cp}=2\times 1.010$. This is the mass of the \cp\ generated
approximate kink, which is within
$1\%$ of the energy of the exact 1-kink solution $\okink$.
Let us note that not only is the procedure we have described very
similar to the Atiyah-Manton procedure but the results 
in the 1-soliton sector have a
remarkable similarity also. Compare the form of the approximate kink
$\cpo$ with the approximate profile function $\aspf$, the values
of the instanton scales at which the approximate soliton energies
are minimized, and the percentage error by which the energy of each approximate
soliton exceeds that of the exact solution.

Now consider the case $n=2$. In order to make the interpretation
of each coordinate as clear as possible we write $W$ in the form
$$W^{-1}={\lambda_1^{-1}\over(x_+-a_1)}+
{e^{i\theta}\lambda_2^{-1}\over(x_+-a_2+ic)}
\eqn\wtwo$$
where $\lambda_1$,$\lambda_2$ are the positive scales of each
instanton and $a_1,a_2,c$ are real and give the position on the 
$x_1$ axis of each instanton and their separation in $x_2$ space
respectively, and $\theta$ gives the relative internal phase
between the two instantons. 

We have that ${\cal M}_2^{cp}=S^1\times{\rm\R}^3\times({\rm\R}^+)^2$.
However, as we shall demonstrate shortly, the manifold ${\cal M}_2^{cp}$ 
is not a good truncation of the 2-kink sector, unless we restrict to
a certain submanifold. Before we begin this discussion let us make
a few simplifications by considering only two-kink
configurations in which the kinks are placed symmetrically about the
origin and have momenta of equal magnitude but opposite sign \ie we
are considering 2-kink solutions that are given by the exact solution
$\tkink$. This corresponds to setting $\lambda_1=\lambda_2=\lambda$ and
$a_1=-a_2=a$; we also choose to set $c=0$ for simplicity. Our first
argument is no more than an intuitive observation as to why the 
value of $\theta$ should be taken to be zero. We then make an explicit
computation to show that the manifold obtained by allowing $\theta$
to take arbitrary values results in an incorrect 
approximation to the sine-Gordon model, since it predicts a
spurious 2-kink bound state.

As noted earlier an important difference between the Skyrme
theory and the sine-Gordon theory is that the solitons of the latter
have no internal degrees of freedom \ie no orientation or phase.
We therefore conjecture that to approximate sine-Gordon solitons
one should, in contrast to the Skyrme case, consider only instantons
which are all in phase. For $n=2$ this means setting $\theta=0$.
We have now restricted to a two dimensional submanifold of
${\cal M}_2^{cp}$ which we denote by 
$\tilde{\cal M}_2^{cp}$. 
In fact $\tilde{\cal M}_2^{cp}={\rm\R}^+\times{\rm\R}^+$, where the
coordinates are $a$ and $\lambda$. Note that $a$ is no longer allowed
to be equal to zero since the two-instanton solution would
then degenerate into a one-instanton solution. Explicitly from
$\tilde{\cal M}_2^{cp}$ we obtain
(after calculating the integral in $\myhol$ by contour integration)
 the 2-kink field
$${\Tenpoint
\eqalign{
&\phi_2^{cp}(x,\lambda,a)=
2\pi x(1+{x^2-a^2+2\lambda^{-2}\over
\sqrt{(x^2-a^2)^2+4x^2\lambda^{-2}}})\times
\cr
&\hskip -0.3cm
{1\over
\sqrt{x^2+a^2+2\lambda^{-2}+2\sqrt{a^2(x^2+\lambda^{-2})+\lambda^{-4}}}
+\sqrt{x^2+a^2+2\lambda^{-2}-2\sqrt{a^2(x^2+\lambda^{-2})+\lambda^{-4}}}}
}}\eqn\cpt
$$
If we compute numerically the energy of this configuration
as a function of $a$ and $\lambda$ we find that it is 
minimized in the limit $a\rightarrow\infty$, in which $\cpt$ becomes
asymptotically two approximate kinks of the form $\cpo$ with positions
$a$ and $-a$. The energy in this limit is therefore minimized when
$\lambda=\lambda_{cp}$ at which it takes the value $E=2M_{cp}$.
$\tilde{\cal M}_2^{cp}$ therefore produces the correct 
qualitative sine-Gordon behaviour of no bound states in the
two-soliton sector. Note that this feature is not automatic 
for our approximate 2-soliton sector since the approximate 1-soliton
does not attain the topological lower bound \ie $M_{cp}>2$.
We will now demonstrate that it is precisely this  qualitative 
\lq\lq no bound states\rq\rq\ 
requirement which is violated by ${\cal M}_2^{cp}$ as a finite
dimensional truncation. In the same way that we defined the 
submanifold $\tilde{\cal M}_2^{cp}$ we define the similar two
dimensional submanifold $\hat{\cal M}_2^{cp}$ but where we set
$\theta=\pi$ rather than the previous case of $\theta=0$. We are
now considering the case in which the two instantons are exactly
out of phase. Note that $\hat{\cal M}_2^{cp}=\R\times\R^+$ since the
position coordinate $a$ is now allowed to take the value zero, since
this 2-instanton solution no longer degenerates into a 1-instanton
solution in the $a\rightarrow 0$ limit now that the
 instantons are not identical
(they have opposite orientation).
Explicitly from $\hat{\cal M}_2^{cp}$ we obtain the two kink field
(where for notational convenience we define $\zeta=\lambda^2x^2-1$)
{\Twelvepoint
$$\hat\phi_2^{cp}(x,\lambda,a)=
\cases{
 {\pi x\sqrt{{\lambda\over\zeta}}}\bigl(
{\sqrt{\zeta}+\lambda a \over \sqrt{\lambda(x^2+a^2)+2a\sqrt{\zeta}}}
+{\sqrt{\zeta}-\lambda a \over \sqrt{\lambda(x^2+a^2)-2a\sqrt{\zeta}}}
\bigr)  
& if $ \zeta \geq 0$\cr
\cr
{\pi xa\sqrt{2\lambda} \over \zeta \sqrt{\lambda^2(x^2+a^2)^2+4a^2\zeta^2}}
\times\cr
\bigl\{ \zeta
\sqrt{\sqrt{\lambda^2(x^2+a^2)^2+4a^2\zeta^2}+\lambda(x^2+a^2)}
\cr
+\lambda a
\sqrt{\sqrt{\lambda^2(x^2+a^2)^2+4a^2\zeta^2}-\lambda(x^2+a^2)}
\bigr\}
& if $\zeta < 0.$
\cr
} \eqn\atapp 
$$}
Of course in the limit $a\rightarrow\infty$ this solution
again becomes two approximate 1-kinks as before, with the energy
minimized at $\lambda=\lambda_{cp}$ and taking the value $2M_{cp}$.
However, the crucial point is that the energy of $\atapp$ is not
minimized by this limit of $a$. Numerically calculating the energy
of $\atapp$
we find it is minimized at $a=3.16$ and $\lambda=0.55$,
at which it has the value $E=4\times 1.003 <2M_{cp}$. The exact values
of these constants are unimportant, the point being that the energy is
minimized at a finite value of $a$. This means that $\hat{\cal M}_2^{cp}$
does not provide a qualitatively good truncation of the 2-soliton
sector since it describes a theory with a 2-kink bound state.
This observation is a warning that care
must be taken when applying the instanton method to the study of 
solitons. If we had been studying a non-integrable soliton theory,
with no exact results to guide us, then the instanton method may
have led us to the incorrect conclusion that a 2-kink bound state
existed in the full theory.
It would appear that the only way to attempt to avoid making
such errors in non-integrable theories is the kind of intuitive argument
given above, based upon relating the instanton
orientation to the soliton orientation (or lack of it).

It is in principal possible to calculate explicitly the 2-kink
field corresponding to the manifold ${\cal M}_2^{cp}$, since 
using contour integration methods it depends on the zeros of a 
quartic polynomial, for which a closed form expression exists.
However, we have already demonstrated that a submanifold of
${\cal M}_2^{cp}$, namely $\hat{\cal M}_2^{cp}$, does not provide a
good truncation and this implies that neither does ${\cal M}_2^{cp}$.
This supports our conjecture that the correct manifold to use
is $\tilde{\cal M}_2^{cp}$ obtained from instantons with the same
orientation. In the rest of this paper we will restrict to this
manifold when discussing the 2-soliton sector generated from \cp\
instantons.

\chapter{Kinks from Yang-Mills Instantons}
As briefly described earlier the Atiyah-Manton procedure involves
taking SU(2) Yang-Mills instantons in \R$^4$ and computing a holonomy.
In the previous section we introduced an analogue of this procedure
which again involves computing  holonomies but this time of 
\cp\ instantons in \R$^2$. There is yet another analogue,
 which we shall now discuss,
where we begin with the same instantons used in the
 Atiyah-Manton scheme.

It is obvious that to obtain sine-Gordon kinks from 
SU(2) Yang-Mills instantons in \R$^4$ one has to do something
other than simply compute a holonomy. The problem is how to 
construct a real-valued field in \R\ from such an instanton.
The solution is suggested by equation $\wtopc$ of the previous 
section. There we saw that computing the holonomy $\myhol$
was  equivalent, in that case, to computing the integral of
the topological charge density over a certain region of space.
For other instanton systems there will be no such equivalence
but it is clear that the second interpretation can be taken as a 
definition to {\sl define} the procedure. In reference \REF\rSW\SW
[$\rSW$] the authors considered such a  procedure by defining
$$\phi (x)=2\pi\int_{-\infty}^{x}dx_1\int_{-\infty}^{+\infty}dx_2
\dots\int_{-\infty}^{+\infty}dx_d\ q_d(x_1,\dots ,x_d)
\eqn\swgen$$
where $q_d$ denotes the topological charge density of any
instanton theory in $d$-dimensional euclidean space. Clearly by
construction if we consider an $n$-instanton solution to the 
gauge theory in question then $\phi(\infty)=2\pi n$ and $\phi(0)=0$
so that $B_{sg}=n$. This procedure becomes the holonomy procedure
of the previous section if we choose $d=2$ and the instanton theory
to be the \cp\ $\sigma$-model. The analysis in [$\rSW$] was 
concerned only with the 1-soliton sector where it was shown that,
for the theories they considered, the
best approximate 1-kink was obtained from four-dimensional SU(2)
 Yang-Mills theory. In this section we will therefore concentrate
on this particular instanton theory. 

Let $A_\mu$ be the $su(2)$-valued Yang-Mills gauge potential
for an instanton in four-dimensional euclidean space with
coordinates $x_\mu$, $\mu=1,2,3,4$. The gauge field is
$F_{\mu\nu}=\partial_\mu A_\nu -\partial_\nu A_\mu +[A_\mu,A_\nu]$.
In terms of these quantities the formula $\swgen$ is
$$\phi(x)={-1\over
16\pi}\int_{-\infty}^xdx_1\int_{-\infty}^{+\infty}
dx_2dx_3dx_4\ 
\epsilon_{\mu\nu\alpha\beta}{\rm Tr}(F^{\mu\nu}F^{\alpha\beta})
\eqn\tchym$$
where $\epsilon_{\mu\nu\alpha\beta}$ is the totally antisymmetric
symbol on four indices.

The moduli space of charge $n$ SU(2) Yang-Mills instantons 
has dimension $8n-3$, but since the above integral is invariant
under shifts of $x_j$ for $j=2,3,4$, this procedure produces
kink fields which depend on a manifold, which we call 
${\cal M}_n^{ym}$, which has dimension $8n-6$.

One could now simply substitute an $n$-instanton solution
into  $\tchym$ and attempt to perform the integration, as 
done in [$\rSW]$ for the $n=1$ case. However, let us first comment
that, as mentioned above, $\tchym$ is to be thought of as a similar 
expression to $\wtopc$. It would be interesting, and
certainly extremely useful from a computational point of view, if 
another form of the relation $\tchym$ could be found which was closer
in spirit to the form $\myhol$; in the sense that an integral over
$x_1$ did not have to be performed. 
In the following we restrict to
instantons obtained from a superpotential
\def\thi{\TH\nextline\CF\nextline\WIL}\REF\rthi\thi\REF\rJNR\JNR
[$\rthi,\rJNR$] (\ie 't Hooft or
 the more general JNR instantons) and  show
that such a relation exists. Note that for the cases $n=1,2$ such
a restriction results in no loss of generality since all instantons
are of this form. It is also possible to produce a similar result
in the general instanton case, using the ADHM
[$\REF\rADHM\ADHM\rADHM$] construction and an expression for the
topological charge density as a double laplacian
[$\REF\rCOFN\COFN\rCOFN$], but we shall not describe this here.

The instantons we consider are therefore given by
$$A_\mu={i\over 2}\sigma_{\mu\nu}\partial^\nu\log\rho\eqn\cfthw$$
where $\sigma_{\mu\nu}=(\epsilon_{4\mu\nu\alpha}+\delta_{\mu\alpha}
\delta_{4\nu}-\delta_{4\mu}\delta_{\nu\alpha})\sigma^\alpha$,
 $\sigma_\alpha$, $\alpha=1,2,3$ are the Pauli matrices
and $\rho(x_\mu)$ is the real-valued superpotential satisfying
the wave equation in euclidean four-space $\waveop\rho =0$.
Then $\tchym$ becomes
$$\phi(x)={-1\over
8\pi}\int_{-\infty}^xdx_1\int_{-\infty}^{+\infty}
dx_2dx_3dx_4\ 
\waveop\ \waveop\ \log\rho
\eqn\tchymb$$
where, as in [$\rJNR$], this integral must be interpreted as 
excluding infinitesimal regions of integration containing the singularities
of $\rho$; this is equivalent to defining 
$\waveop\ \waveop\log(x_\mu-X_\mu)^2$ to be zero for
 any constant $X_\mu$ in \R$^4$.
By converting total derivative terms to boundary terms we
write $\tchymb$ in its final form
$$\phi(x)=n({\rm mod2})\pi
-{1\over
8\pi}\int_{-\infty}^{+\infty}
dx_2dx_3dx_4\ 
\partial_1^3 \log\rho\eval_{x_1=x}
\eqn\tchymc$$
where we have succeeded in removing the integral over $x_1$.

Take the case $n=1$, where we may write (using the above mentioned
shift invariances)
$$\rho=1+{\lambda^{-2}\over (x_\mu-a\delta_{1\mu})^2}.\eqn\ymoi$$
Here $\lambda$ is the positive scale of the instanton and 
$a$ is real, giving the position of the instanton on the $x_1$ axis.
${\cal M}_1^{ym}=\R^+\times\R$\ and corresponds to the explicit
1-kink field (obtained from integrating $\tchymc$)
$$\eqalign{\phi_1^{ym}(x,\lambda,a)&
=\pi+{\pi\over 2}\partial_x^2\{(x-a)\sqrt{\lambda^{-2}+(x-a)^2}\ \}
\cr
&=\pi(1+{\lambda(x-a)\over\sqrt{1+\lambda^2(x-a)^2}}
+{{1\over 2}\lambda(x-a)\over (1+\lambda^2(x-a)^2)^{3\over 2}})} 
\eqn\ymok$$
The final expression is the form in which it was obtained in
[$\rSW$] by calculating explicitly the integral $\tchym$.
The energy of $\ymok$, independent of the kink position $a$,
is minimized when $\lambda=\lambda_{ym}=0.419$ at which
it takes the value $E=M_{ym}=2\times 1.0002$. This is the mass of
the Yang-Mills generated approximate kink, which exceeds that of the
exact solution $\okink$ by around $0.02\%$. It is therefore
a remarkably accurate approximation.

We now consider the $n=2$ sector, and following the remarks
of the previous section, consider instantons which have 
no relative orientation. We have not performed an explicit 
calculation to see whether the \lq\lq no bound states\rq\rq\
requirement is violated for Yang-Mills instantons with a relative
orientation. Note that since $(M_{ym}-2)$\ is two orders of
magnitude smaller than
$(M_{cp}-2)$\ the window for such a possible violation is
greatly reduced. Accurate numerical work is required to
settle this issue. We also make the same symmetric kink simplifications
that we made in the previous section, and consider the submanifold
$\tilde{\cal M}_2^{ym}=\R^+\times\R^+$\ corresponding to the
instanton superpotential
$$\rho=1+
{\lambda^{-2}\over (x_\mu-a\delta_{1\mu})^2}
+{\lambda^{-2}\over (x_\mu+a\delta_{1\mu})^2}
\eqn\ymti$$
This gives the 2-kink field
$$\eqalign{&\phi_2^{ym}(x,\lambda,a)
\cr 
&=\pi\partial_x^2\{
{x(x^2+3a^2+\lambda^{-2}+\sqrt{(x^2-a^2)^2+2\lambda^{-2}(x^2+a^2)}\ )
\over
\sqrt{x^2+a^2+\lambda^{-2}+\sqrt{4x^2a^2+\lambda^{-4}}}
+\sqrt{x^2+a^2+\lambda^{-2}-\sqrt{4x^2a^2+\lambda^{-4}}}}
\}}
\eqn\ymtk$$
Note that upon performing the double differential in the above
expression the result will be quite lengthy and would have been
tedious to calculate from the expression $\tchym$ directly.

The energy of $\ymtk$ is minimized in the limit $a\rightarrow\infty$,
when the field becomes two approximate kinks of the form $\ymok$
with positions $a$ and $-a$, so that the energy is minimized when
$\lambda =\lambda_{ym}$ at which it is equal to $2M_{ym}$.

\chapter{Dynamics on ${\cal M}_n^{cp}$ and ${\cal M}_n^{ym}$}
In this section we compute the approximate soliton dynamics, in the
one and two-soliton sectors, obtained from the finite dimensional
truncations constructed in the previous two sections.

We first consider ${\cal M}_1^{cp}$, with parameters $a$ and
$\lambda$. As described in the introduction, time dependence is
introduced by allowing these parameters to be functions of the
\lq\lq physical time\rq\rq\ $t$ \ie $a(t)$ and $\lambda(t)$.
A lagrangian for this system is then obtained by restricting
the field theory lagrangian to the configuration space described
by ${\cal M}_1^{cp}$ \ie the lagrangian is obtained by substituting
the ansatz $\cpo$ into the lagrangian density $\lagsg$ and
integrating over $x$. Before we give the result it is convenient
to introduce the normalized scale $\mu=\lambda/\lambda_{cp}$,
so that the minimum energy of a static configuration is obtained 
at the value $\mu=1$. The result for the lagrangian is
$${\cal L}_1^{cp}={M_{cp}\over 2}(\mu \dot a^2 
+k_{cp}{\dot\mu ^2\over \mu^3}-\mu-{1\over \mu})
\eqn\lcpo$$
where dot denotes differentiation with respect to time $t$
and we have introduced the constant $k_{cp}={1\over 3\lambda_{cp}^2}
\approx 0.69$. It is easy to see that the equations which follow from
$\lcpo$ are 
$$\eqalign{
&\ddot a\mu+\dot a\dot\mu =0\cr
&2k_{cp}\ddot\mu\mu-3k_{cp}\dot\mu^2+\mu^4(1-\dot a^2)-\mu^2=0
}\eqn\fdoseom$$
and have a one parameter family of solutions (independent of $k_{cp}$) given by
$$a=ut,\hskip 1cm\mu=\gamma\eqn\fdoss$$
for $-1<u<1$ constant, where $\gamma$ is the
 Lorentz factor $(1-u^2)^{-{1\over2}}$.
The energy of such a solution is $\gamma M_{cp}$. 
Hence we obtain a simple Lorentz boost of the static approximate kink.
Note that to be able to obtain an exact Lorentz boost from the
instanton scale is a feature unique to the approximation of
solitons in one dimensional space.

If we now consider ${\cal M}_1^{ym}$, by performing the same procedure
with the ansatz $\ymok$ we obtain the same results as above 
after a labelling replacement $cp\rightarrow ym$, where the only
quantity we have yet to define is $k_{ym}={1\over 7\lambda_{ym}^2}
\approx 0.81$. It is interesting to compare these two results with
that obtained by taking the exact kink solution $\okink$ as an ansatz
and introducing a scale through the replacement $x\rightarrow\mu x$.
We then get a similar result with the obvious notation $M_{exact}=2$
and $k_{exact}={\pi^2\over 12}\approx 0.82$. This demonstrates that 
the coupling of translational and scale degrees of freedom is very
similar for the approximate kinks as it is for the exact kink;
compare $k_{cp}$ with $k_{exact}$ and particularly $k_{ym}$
 with $k_{exact}$.

Let us now turn to the 2-soliton sector and consider $\tilde{\cal M}_2^{cp}$.
We are going to make one further simplification  (it turns out that
we obtain almost identical results with or without this
simplification). In the limit of large $a$ (\ie well-separated kinks)
a 2-kink configuration becomes two 1-kink configurations so that the
lagrangian becomes ${\cal L}_2^{cp}\approx 2{\cal L}_1^{cp}$.
The solution in this region will therefore be two free, Lorentz
boosted, kinks with $\dot a=-u$, $\lambda=\gamma\lambda_{cp}$,
corresponding to studying kink scattering with kink velocity
$0<u<1$; compare the exact solution $\tkink$. Our simplification
is to assume that the scale changes little during the evolution
so that we can fix it at this value \ie we set $\dot\lambda=0$, and
truncate to a system with only one degree of freedom. (This is in
fact a  good approximation. We have studied the two-dimensional
system, which arises by allowing  $\lambda(t)$, numerically using
a Runge-Kutta evolution and find that the variation of $\lambda$
from the constant value $\lambda=\gamma\lambda_{cp}$ is so small
as to have a negligible effect on our results).

Substituting the ansatz $\phi_2^{cp}(x,\gamma\lambda_{cp},a(t))$
into $\lagsg$ and integrating over $x$ gives
$${\cal L}_2^{cp}={1\over 2}g_{cp}(a)\dot a^2 -V_{cp}(a)
\eqn\cplt$$
where the metric  $g_{cp}$ and the potential 
$V_{cp}$ are given by
$$\eqalign{
&g_{cp}(a)={1\over 2}\int_0^{\infty}dx\ ({\partial\phi_2^{cp}
\over \partial a}(x,a,\gamma\lambda_{cp}))^2 \cr
&V_{cp}(a)=\int_0^{\infty}dx\ {1\over 4} 
({\partial\phi_2^{cp}
\over \partial x}(x,a,\gamma\lambda_{cp}))^2
+{1\over 2}(1-\cos\phi_2^{cp}(x,a,\gamma\lambda_{cp})).}
\eqn\gv$$
Of course, in the Yang-Mills case we have exactly the same formalism,
again simply replacing $cp$ by $ym$ so that we are then using the 
function $\phi_2^{ym}(x,\gamma\lambda_{ym},a(t))$ {\sl etc}.
In Fig. 1. we plot the metric and potential functions 
$g_{cp},V_{cp},g_{ym},V_{ym}$ obtained by integrating the above
expressions numerically.\endpage

\def\capa{The metric and potential functions $g_{cp},V_{cp}$
(continuous curves) and $g_{ym},V_{ym}$ (dashed curves).
In each case the potential curves are the monotonic decreasing 
functions of a.}

\epsfxsize=8cm \epsfbox{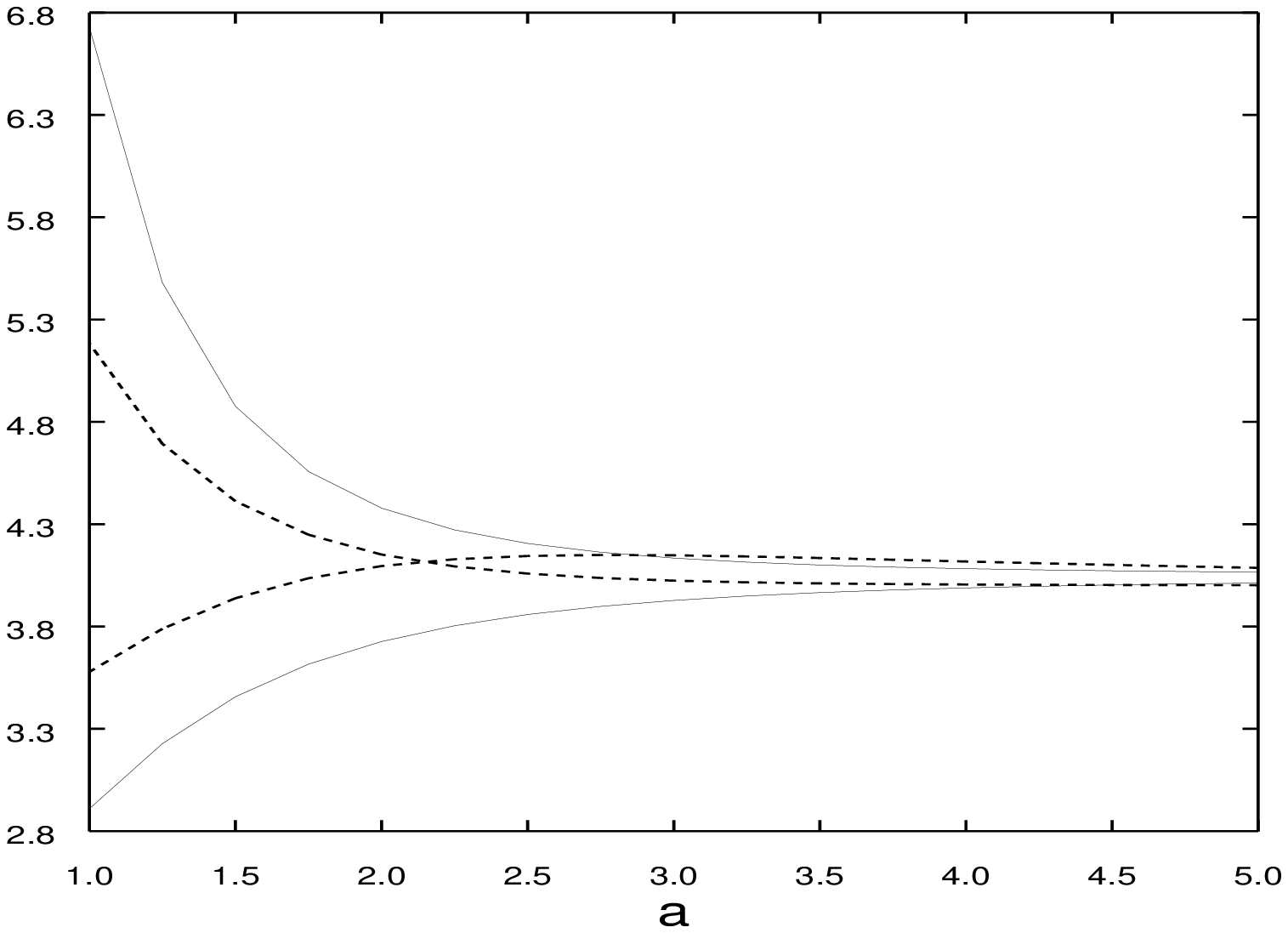} \item{Fig\,1:} {\capa}

To produce the  plots of Fig 1. we have set $\gamma=1$. The range
of $a$ reflects the values which are relevant in the scattering 
processes we shall discuss below. It is easy to see that since 
$\phi_2^{cp}$ is a symmetric function of $a$ then $g_{cp}\rightarrow
0$ as $a\rightarrow 0$. Also it can be shown that the slope of
$\phi_2^{cp}$ at the origin behaves like ${1\over a^2}$, so that the
potential grows without bound as $a\rightarrow 0$. Similar remarks
apply for $\phi_2^{ym}$.
From $\cplt$ the equations of motion are
$$g_{cp}\ddot a +{1\over 2}{dg_{cp}\over da}\dot a^2 + {dV_{cp}\over da}=0
\eqn\cptkeom$$
with initial conditions $a(t=0)=a_0$, $\dot a(t=0)=-u$.

The solution can be obtained implicitly by quadrature as
$$t(a)=\int_a^{a_0}\sqrt{{g_{cp}(\alpha)\over 2(E-V_{cp}(\alpha))}}\ d\alpha
\eqn\solnquad$$
where $E={1\over 2}g_{cp}(a_0)u^2+V_{cp}(a_0)$ is the total energy.
This solution is valid upto $t=t_1$, where $t_1$ is the 
turning time $t_1=t(a_1)$ and $a_1$ is the turning point
$V_{cp}(a_1)=E$. 
The position for $t>t_1$ can be determined from the symmetric 
property of the motion $a(t-t_1)=a(t_1-t)$. The approximate 2-kink
phase shift is given by
$$\delta_{cp}=2(a_0-ut_1)\eqn\apps$$
in the limit in which $a_0\rightarrow\infty$ with $t_1$ determined as
above.
 
We now compare the results of our finite dimensional truncations
with the exact results given in section 2. First note that to compare
with the exact 2-kink solution $\tkink$ it is necessary to shift the
time coordinate $t$ in this solution to 
$t-u^{-1}({\log(u^{-1})\over\gamma}-a_0)$, so that the kinks have
positions $\pm a_0$ at $t=0$. In Fig. 2a. we plot the function
$\phi_2^{cp}(x,\gamma\lambda_{cp},a(t))$ (continuous curve)
and the exact solution $\tkink$ (dashed curve) for a velocity
$u=0.5$, at various times. We have taken $a_0=10$ and all 
required integrals are performed numerically. Fig. 2b. is as
Fig. 2a. but for the 
Yang-Mills function 
$\phi_2^{ym}(x,\gamma\lambda_{ym},a(t))$
with the time evolution of $a$ determined by
${\cal L}_2^{ym}$.
\endpage
\def\capb{$\phi_2^{cp}$ (continuous curve)
and the exact solution (dashed curve), at increasing times.}
\def\capc{As Fig 2a. but for the approximation $\phi_2^{ym}$.}

\pageinsert\phantom{xx}\vskip 7.5cm
\epsfxsize=7cm\epsfbox{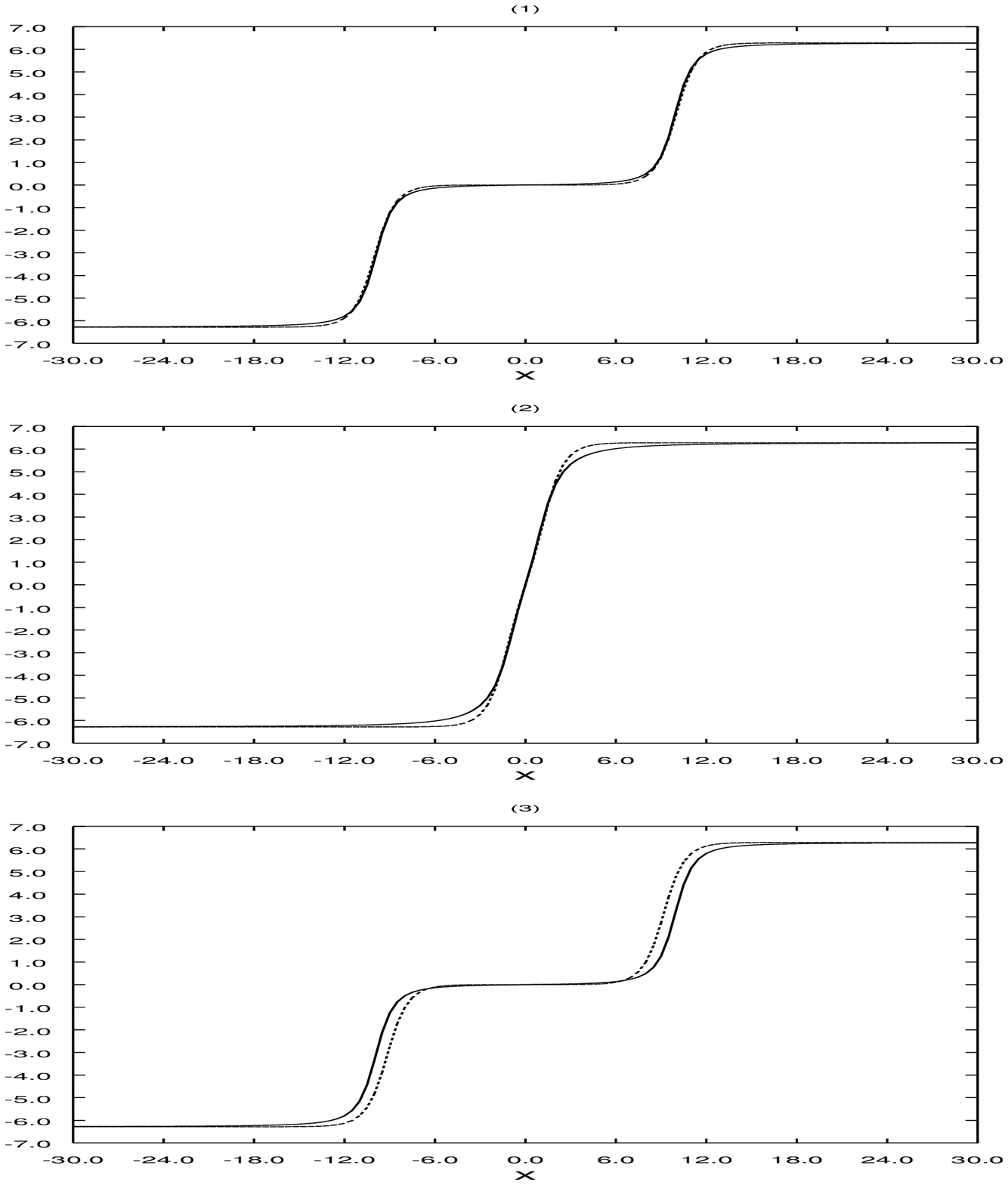}\item{Fig\,2a:}{\capb}\endinsert
\pageinsert\phantom{xx}\vskip 7.5cm
\epsfxsize=7cm\epsfbox{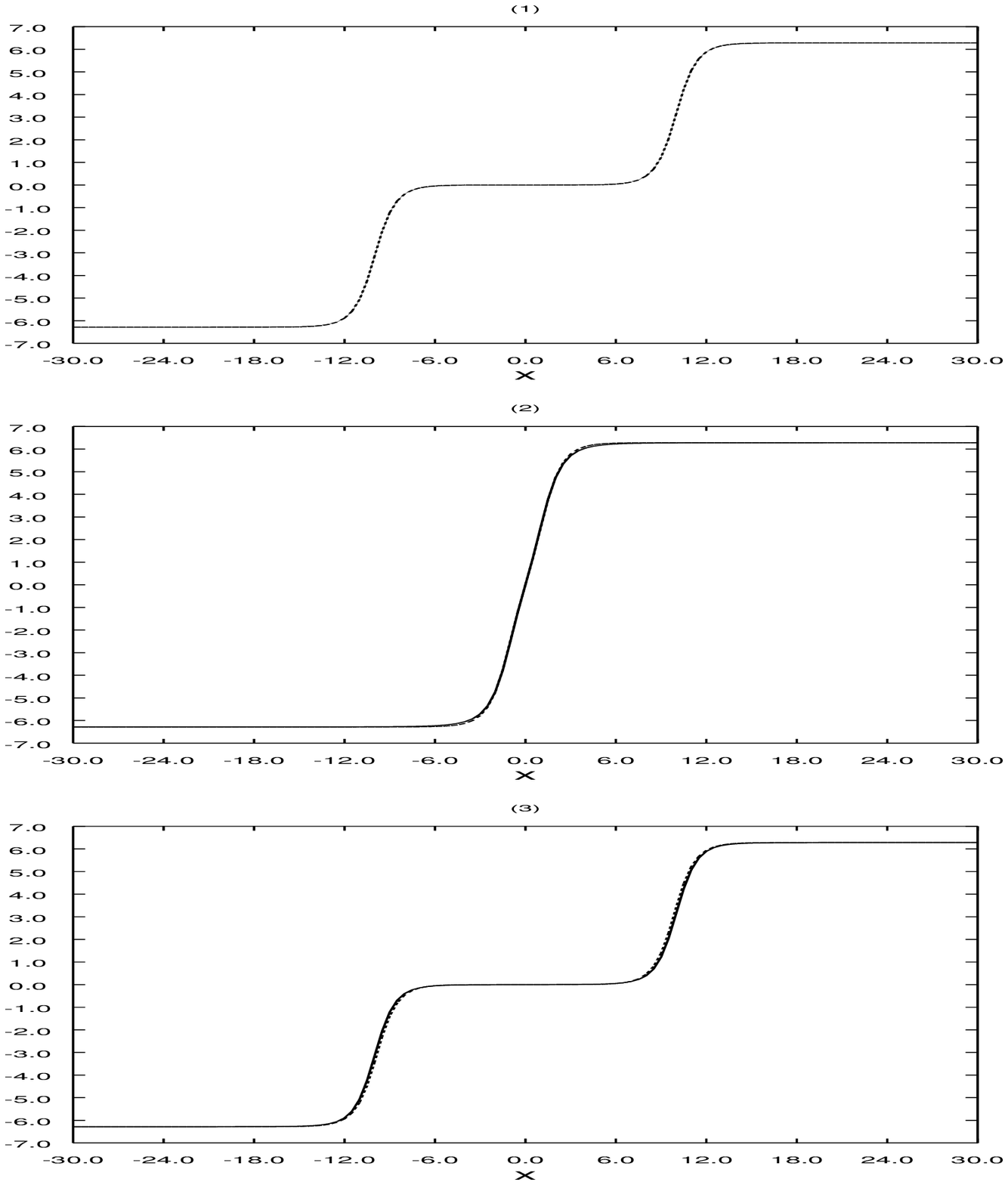}\item{Fig\,2b:}{\capc}\endinsert

Fig 2a.(1) shows the initial ($t=0$) configuration, and it can be
seen that it is very close to the initial exact solution.
The kinks then collide, and  Fig 2a.(2) shows the configurations
at the time for which the separation $a$ is a minimum.
Note that since we are considering the scattering of identical
kinks it is impossible to distinguish forward from backward 
scattering. Our instanton moduli coordinates make backward
scattering the most natural interpretation (since $a\in\R^+$)
but we interpret the results in the more usual forward scattering 
manner in order to
compare with the exact results. We can see that even when the
kinks have lost their individual identities the approximation is still
quite good, although it is clear that there are two regions in which
the error is now appreciable. Fig 2a.(3) corresponds to the time 
for which the position $a$ has returned to its initial value.
The error here is quite clear, with the approximation giving
a phase shift in excess of the exact result.

We now turn to Fig 2b. These figures correspond to those of Fig 2a.
but now for the Yang-Mills generated approximation. Again two
curves are graphed in each window, but the approximation is so close
to the exact solution that it appears there is only a single graph!
This approximation is clearly extremely accurate and much the
superior of the two schemes. This is to be expected since the 
one-kink field $\phi_2^{ym}$ is a much better approximation than
the one-kink field $\phi_2^{cp}$. 

In Fig. 3. we plot the exact phase shift $\delta$ (continuous curve),
the \cp\ generated approximation $\delta_{cp}$ (stars) and the
Yang-Mills generated approximation $\delta_{ym}$ (circles) for 
velocities in the range $0.1\le u\le 0.9$. This figure displays
the main result of this paper, showing how accurately the 
instanton truncations approximate the infinite dimensional field
theory dynamics. Although the $cp$ approximation gives qualitatively
correct results the errors are quite large. This may seem surprising
when one considers the accuracy of the
one-kink field $\phi_1^{cp}$; recall its energy is only 1\% larger
than that of the exact one-kink. However, an important point to bear
in mind is that the approximation decays power-like, as opposed to the
exponential decay of the exact one-kink. This would therefore
suggest that we would be over optimistic to hope to obtain
results of too great an accuracy. This is where the $ym$ approximation
yields surprising results. It too generates kinks with a power-like
tail behaviour, but from Fig. 3. we see that the accuracy is
remarkable, for the whole range of velocities. So it would seem that
the exponential localization of sine-Gordon solitons  is not a major
factor in influencing their interaction. This is an interesting point
which requires further investigation and is of  relevance to higher
dimensional systems where solitons are often localized
in a power-like way {\sl eg} lumps in planar sigma models.

\def\capd{The exact phase shift $\delta$ (continuous curve),
and the approximations $\delta_{cp}$ (stars), and
$\delta_{ym}$ (circles) for velocities $0.1\le u\le 0.9$.}

\epsfxsize=8cm \epsfbox{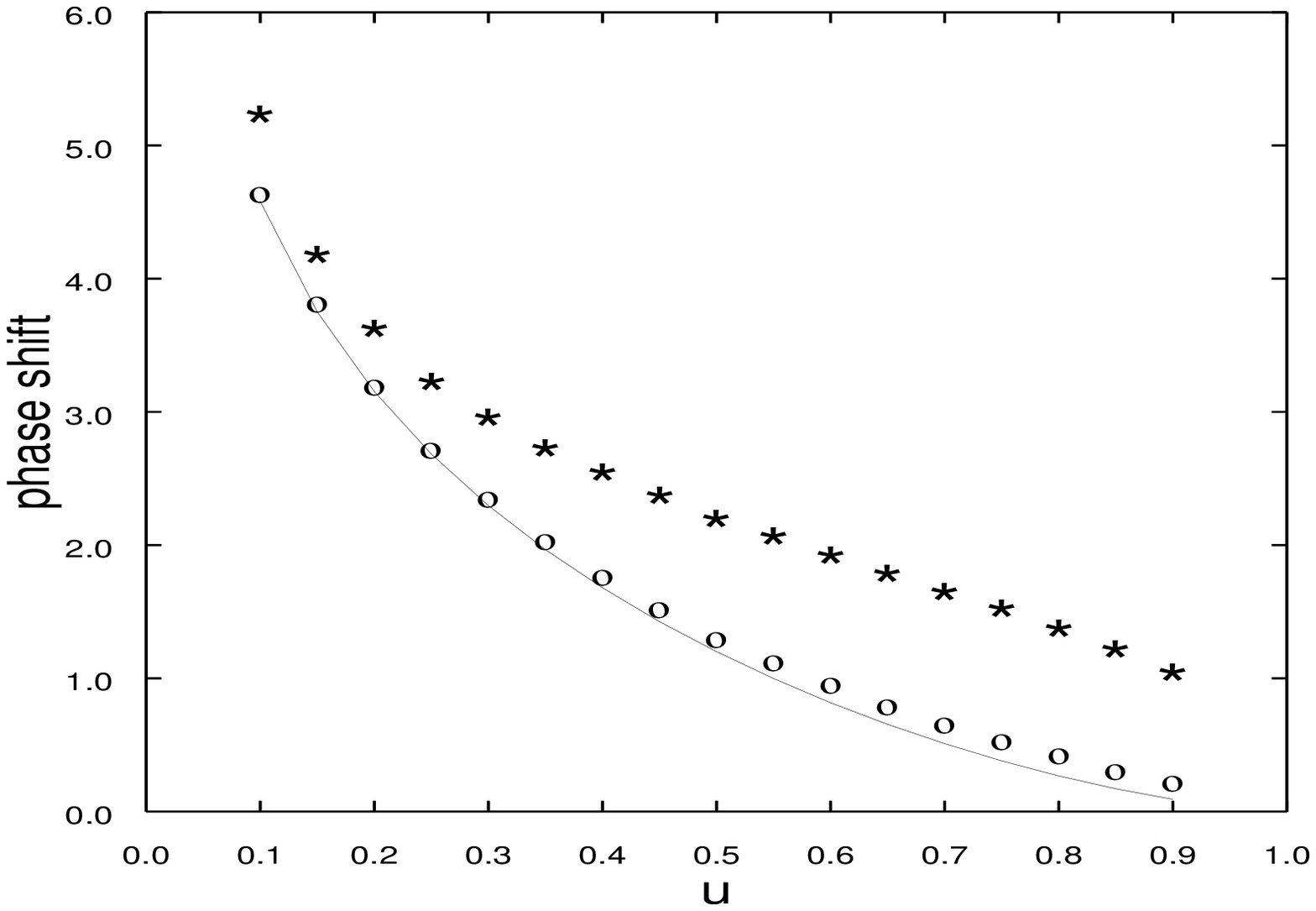} \item{Fig\,3:}{\capd}

\endpage
\chapter{Conclusion}
This paper has dealt with some simpler analogues of the Atiyah-Manton
construction of Skyrmions from instantons. We have shown how their
procedure may be adapted to study other topological solitons, and
furthermore have demonstrated the validity of the scheme in a case
where exact results are available for comparison. 
The discovery that constraints may have to be imposed upon the 
instanton orientations, in order to correctly reproduce the
qualitative bound states behaviour of the soliton theory, is 
unexpected. It is clearly an important issue to consider when
applying the scheme to non-integrable soliton theories.
For the particular case of the Skyrme model the intuitive 
argument described earlier suggests that no constraint is 
required, since the orientation of the Yang-Mills instanton
produces the required isospin of the Skyrmion.

The results in this paper are  encouraging
for the use of the instanton method in approximating the dynamics
of Skyrmions. As mentioned earlier, there are as yet no complete
results on the instanton truncation of Skyrmion dynamics, mainly
due to the heavy computations this involves, and some technical
\def\HOS{\HGOA\nextline\HOA}
difficulties in computing the non-abelian holonomy. Using numerical
methods \REF\rHOS\HOS[$\rHOS$]
 there are some results regarding the potential
function on the finite dimensional two-Skyrmion manifold, and perhaps
a full numerical scheme could also be used to construct the metric
function and solve the resulting ordinary differential equations
which result. 

In this paper we have dealt with only the classical dynamics of
solitons. It is also possible to study quantum scattering by
a quantization of the finite dimensional system obtained from
instanton moduli. A finite dimensional truncation of the sine-Gordon
model (obtained from patched kinks) was quantized in
\REF\rPMSF\PMSF[$\rPMSF$] and found to compare well with the 
known exact results in the weak coupling limit. A similar analysis
carries over to the finite dimensional manifolds constructed
in this paper and results of a similar quality can be obtained.

It appears that the soliton from instanton construction is a useful 
approximation technique, the wide applicability of which is only 
just beginning to emerge. It is already known that this scheme can
be applied to other kink systems, such as the double sine-Gordon
equation [$\rSW$], to produce accurate approximate static kinks, and it 
produces excellent results when applied to  $\phi^{2n}$ field
theory \REF\rPMSH\PMSH[$\rPMSH$]. No doubt the method will prove
to be a useful tool as other areas of application come to light.

Finally, let us make a few remarks about Skyrmions which arise from
the work in this paper. One point to note is that our earlier
comments about the power-like decay properties of the approximate kinks,
as compared to the exponential decay of the exact kink, also have
relevance in the Skyrme context. The instanton generated 
profile function $\aspf$ decays like  ${1\over r^2}$ for large $r$,
as does the (numerical) exact solution. However, if a pion mass
term is added then the exact solution now has an exponential decay.
The results of this paper would seem to suggest that this is not 
a serious defect if one wished to use the Atiyah-Manton scheme to study
Skyrmion dynamics even in the massive pion case. A further observation
is based on the similarity of the Skyrmion profile
functions $\aspf$ and the kink function $\cpo$. In fact, if one
considers a $\CP^1$ anti-instanton (rather than an instanton) positioned
at the origin and makes the replacement $x\rightarrow r$ then one
obtains exactly the profile function $\aspf$. This identification
prompted the use of a sine-Gordon anti-kink field as a trial
function for the Skyrme profile function \REF\rPMSE\PMSE[$\rPMSE$]
and slightly reduced the energy as compared to $\aspf$. Now note that
we can use
a similar identification with a Yang-Mills generated anti-kink, to
propose a modification of the Atiyah-Manton profile function $\aspf$
to 
$$f(r)=\pi(1-{\lambda r\over\sqrt{1+\lambda^2 r^2}}
-{{1\over 2}\lambda r\over(1+\lambda^2 r^2)^{3\over 2}}).\eqn\newpf$$
This function is the same as $\aspf$ but with an additional term
added which does not affect the boundary conditions. As described
earlier $\aspf$ gives a Skyrme field with energy about 1\% above
the (numerical) exact solution, whereas $\newpf$ reduces this excess
to around ${1\over 3}$\%. If one replaces
the ${1\over 2}$ coeffecient of the final term in $\newpf$
with an arbitrary constant $c$, 
 then the
energy is minimized for $c\approx 0.4$ at which the excess energy is
less than ${1\over 10}\%$. By introducing independent scales for the two
terms in $\newpf$ one could perhaps reduce the energy still further.

\ack
I thank the EPSRC for a research fellowship.

\endpage
\refout
\end